\documentclass[a4paper,12pt]{article}
\pdfoutput=1 

\usepackage{jheppub} 

\usepackage[T1]{fontenc} 

\usepackage{amsmath}
\usepackage{graphicx,slashed,xcolor,multirow,bbold,mathtools,sidecap,tikz,bm,enumitem,booktabs,array}

\usepackage[normalem]{ulem}

\usepackage{tikz}
\usetikzlibrary{positioning,arrows}
\usetikzlibrary{decorations.pathmorphing}
\usetikzlibrary{decorations.markings}

\usepackage{subcaption}

\usepackage[compat=1.1.0]{tikz-feynman}
\newcommand{\beq}{\begin{equation}}
\newcommand{\eeq}{\end{equation}}
\newcommand{\bea}{\begin{eqnarray}}
\newcommand{\eea}{\end{eqnarray}}
\newcommand{\barr}{\begin{array}}
\newcommand{\earr}{\end{array}}

\def \st1{\widetilde t_1}
\def \mst1{m_{\st1}}
\def \sbot1{\widetilde b_1}

\def \lspone{\widetilde\chi_1^0}
\def \mlspone{m_{\lspone}}
\def \lsptwo{\widetilde\chi_2^0}
\def \mlsptwo{m_{\lsptwo}}

\def \lspthree{\widetilde\chi_3^0}
\def \lspfour{\widetilde\chi_4^0}

\def\chonepm{\widetilde{\chi}_1^{\pm}}
\def\chtwopm{\widetilde{\chi}_2^{\pm}}

\def\mchonepm{m_{\chonepm}}

\def\bea{\begin{eqnarray}}
	\def\eea{\end{eqnarray}}

\def \met{\rm E{\!\!\!/}_T}

\usepackage{color}
\usepackage{colortbl}

\long\def\/*#1*/{}
\usepackage{color}
 \usepackage[normalem]{ulem}
\definecolor{darkgreen}{cmyk}{1,0,1,0.4}
\definecolor{darkred}{cmyk}{0,1,1,0.4}

\definecolor{lime}{HTML}{A6CE39}
\DeclareRobustCommand{\orcidicon}{\hspace{-2.1mm}
\begin{tikzpicture}
\draw[lime, fill=lime] (0,0) circle [radius=0.13] node[white] {{\fontfamily{qag}\selectfont \tiny \,ID}}; \draw[white, fill=white] (-0.0525,0.095) circle [radius=0.007]; 
\end{tikzpicture} \hspace{-3.2mm} }
\foreach \x in {A, ..., Z}{\expandafter\xdef\csname orcid\x\endcsname{\noexpand\href{https://orcid.org/\csname orcidauthor\x\endcsname} {\noexpand\orcidicon}}}

\title{Probing sub-TeV Higgsinos aided by a ML-based top tagger in the context of Trilinear RPV SUSY}

\author[a]{Rajneil Baruah\orcidB{}}
\author[b]{Arghya Choudhury\orcidC{}}
\author[c,d]{Kirtiman Ghosh\orcidG{}}
\author[a]{Subhadeep Mondal\orcidM{}}
\author[c,d]{Rameswar Sahu\orcidA{}}

\affiliation[a]{\footnotesize Department of Physics, SEAS, Bennett University, Greater Noida, Uttar Pradesh 201310, India}
\affiliation[b]{\footnotesize Department of Physics, Indian Institute of Technology Patna,  Bihar 801106, India}
\affiliation[c]{\footnotesize Institute of Physics, Bhubaneswar, Sachivalaya Marg, Sainik School Post, Bhubaneswar 751005, India}
\affiliation[d]{\footnotesize Homi Bhabha National Institute, Training School Complex, Anushakti Nagar, Mumbai 400094, India}

\emailAdd{rajneilb.physics@gmail.com}
\emailAdd{arghya@iitp.ac.in}
\emailAdd{kirti.gh@gmail.com}
\emailAdd{subhadeep.mondal@bennett.edu.in}
\emailAdd{rameswar.s@iopb.res.in}

\abstract{Probing higgsinos remains a challenge at the LHC owing to their small production cross-sections and the complexity of the decay modes of the nearly mass degenerate higgsino states. The existing limits on higgsino mass are much weaker compared to its bino and wino counterparts. This leaves a large chunk of sub-TeV supersymmetric parameter space unexplored so far. In this work, we explore the possibility of probing higgsino masses in the 400 - 1000 GeV range. We consider a simplified supersymmetric scenario where R-Parity is violated through a baryon number violating trilinear coupling. We adopt a machine learning-based top tagger to tag the boosted top jets originating from higgsinos, and for our collider analysis, we use a BDT classifier to discriminate signal over SM backgrounds. We construct two signal regions characterized by at least one top jet and different multiplicities of $b$-jets and light jets. 
	Combining the statistical significance obtained from the two signal regions, we show that higgsino mass as high as  925 GeV can be probed at the high luminosity LHC.}


\begin{document} 

\tikzset{
  vector/.style={decorate, decoration={snake,amplitude=.4mm,segment length=2mm,post length=1mm}, draw},
  tes/.style={draw=black,postaction={decorate},
    decoration={snake,markings,mark=at position .55 with {\arrow[draw=black]{>}}}},
  provector/.style={decorate, decoration={snake,amplitude=2.5pt}, draw},
  antivector/.style={decorate, decoration={snake,amplitude=-2.5pt}, draw},
  fermion/.style={draw=black, postaction={decorate},decoration={markings,mark=at position .55 with {\arrow[draw=blue]{>}}}},
  fermionbar/.style={draw=black, postaction={decorate},
    decoration={markings,mark=at position .55 with {\arrow[draw=black]{<}}}},
  fermionnoarrow/.style={draw=black},
  scalar/.style={dashed,draw=black, postaction={decorate},decoration={markings,mark=at position .55 with {\arrow[draw=blue]{>}}}},
  scalarbar/.style={dashed,draw=black, postaction={decorate},decoration={marking,mark=at position .55 with {\arrow[draw=black]{<}}}},
  scalarnoarrow/.style={dashed,draw=black},
  electron/.style={draw=black, postaction={decorate},decoration={markings,mark=at position .55 with {\arrow[draw=black]{>}}}},
  bigvector/.style={decorate, decoration={snake,amplitude=4pt}, draw},
  particle/.style={thick,draw=blue, postaction={decorate},
    decoration={markings,mark=at position .5 with {\arrow[blue]{triangle 45}}}},
  gluon/.style={decorate, draw=black,
    decoration={coil,aspect=0.3,segment length=3pt,amplitude=3pt}}
}

\maketitle
\flushbottom

\section{Introduction}
\label{sec:intro}
The approach of the Large Hadron Collider (LHC) towards its high luminosity era opens up new exciting possibilities, especially for new physics scenarios with smaller cross-sections. These scenarios largely remain unexplored owing to low statistics at the LHC. 
Although a large portion of the sub-TeV parameter space of the Minimal Supersymmetric Standard Model (MSSM) has already been ruled out, there are pockets that are yet to be explored.
In the R-parity-conserving (RPC) MSSM scenario, R-parity is introduced in an ad hoc manner to prevent proton decay, resulting in a stable lightest supersymmetric particle (LSP) that can serve as a promising candidate for dark matter~\cite{Martin:1997ns}. However, no fundamental theoretical principles forbid the violation of R-parity.  The R-parity-violating (RPV) MSSM~\cite{Barbier:2004ez, Dreiner:1997uz, Choudhury:2024ggy} scenarios are  well-motivated
and can explain the light neutrino masses and mixings~\cite{Grossman:2003gq, Choudhury:2023lbp, Choudhury:2024yxd}, which the RPC MSSM cannot.
The value of the $\mu$-parameter which is the higgsino mass parameter, is crucial in determining the electroweak scale in supersymmetry (SUSY). This makes sub-TeV higgsino of particular interest from the perspective of natural SUSY. 
After electroweak symmetry breaking, the mixture of Higgsinos and electroweak Gauginos (wino and bino) produces the chargino ($\chonepm, \chtwopm$) and 
the neutralino ($\lspone, \lsptwo, \lspthree, \lspfour$) mass eigenstates, which are known as electroweakinos. 
In the limit, $|\mu|$ $<< |M_1|$ (bino mass), $|M_2|$ (wino mass), the two lighter neutralinos ($\lspone, \lsptwo$) and the lighter chargino ($\chonepm$)  are all mostly higgsinos with approximately the same mass determined by  
the higgsino mass parameter $\mu$. In this article, we shall refer to these 
mass degenerate $\lspone, \lsptwo$, and $\chonepm$ collectively as Higgsinos.
Higgsinos typically have a smaller production cross-section, and as a result, limits on pure higgsino masses are much weaker compared to the bino and wino LSP scenarios~\cite{CMS:2024gyw, ATLAS:2024qxh}.

Both the ATLAS and CMS collaborations have searched for electroweakinos~\cite{CMS:2024gyw, ATLAS:2024qxh} across various final states, including 
trilepton~\cite{ATLAS:2021moa}, soft dilepton~\cite{ATLAS:2019lng}, low momentum displaced tracks~\cite{ATLAS:2024umc} 
\footnote{Several phenomenological groups have also analyzed both RPC and RPV SUSY scenarios with light electroweakinos in the context of dark matter, muon (g-2), interpretation of LHC limits etc.\cite{Ellis:2024ijt, Choudhury:2016lku, Chakraborti:2014gea, Chakraborti:2021bmv, Chakraborti:2015mra, Endo:2021zal, Choudhury:2017fuu, Kowalska:2018toh, Choudhury:2024crp, Chakraborti:2024pdn,  Barman:2020azo, Dreiner:2023bvs, Choudhury:2023eje, Chakraborty:2015bsk}.}.     
Using the LHC Run-II data, they have excluded nearly degenerate Higgsinos with masses up to $\sim$ 170-210 GeV~\cite{CMS:2024gyw, ATLAS:2024qxh} depending 
on the mass gap among the Higgsinos in the RPC SUSY frameworks.
The limit is more stringent in general gauge mediated SUSY scenario and reaches close to $\sim$ 950 GeV when NLSP higgsino decays into gravitino LSP \cite{ATLAS:2024zzz,ATLAS:2024tqe,CMS:2022vpy}.
When R-parity is broken, the higgsino can decay into SM particles directly or via the lightest SUSY particle. Under such scenarios,  the limits on higgsino masses alter \cite{ATLAS:2021fbt, ATLAS:2023lfr} and we need different dedicated search strategies for Higgsinos. The ATLAS collaboration has explored Higgsino production in the RPV SUSY models with bilinear \cite{ATLAS:2023lfr} and  
UDD type coupling \cite{ATLAS:2021fbt, ATLAS:2023lfr}, and excluded Higgsinos masses up to 440 GeV and 320 GeV, respectively. Even with R-parity violation, a large portion of the sub-TeV higgsino mass parameter space remains to be probed.

In this work, we focus on only one type of R-parity breaking term that violates baryon number, namely, $\lambda^{\prime\prime}_{UDD}$. 
It may be noted that $\lambda^{\prime\prime}_{323}$ or $\lambda^{\prime\prime}_{tsb}$ coupling involves the highest possible third-generation quarks and no first-generation quark. 
The $\lambda^{\prime\prime}_{323}$ becomes the largest coupling under minimal flavor violation (MFV)
hypothesis \cite{Csaki:2011ge}. 
The higgsino pair is directly produced via R-parity conserving coupling and spontaneously decays via R-parity violating coupling $\lambda^{\prime\prime}_{323}$. This scenario has recently been explored by the ATLAS collaboration \cite{ATLAS:2021fbt, ATLAS:2023lfr} in multiple signal regions with varied lepton multiplicity. However, the reach on the higgsino mass is restricted to  320 GeV from the LHC run-II data~\cite{ATLAS:2021fbt}. We further probe the scenario with sufficiently heavy Higgsino masses, such that the top quarks produced from Higgsino decays are adequately boosted. In this scenario, one can aim to tag the hadronically decaying top quark as a whole, which can reduce the SM background effectively.

The top quark has only one dominant decay mode, that is, into a $W$-boson and a bottom quark. In traditional LHC searches, the leptonic decay channel of the top quark is often preferred due to the distinct lepton and missing energy signatures that help suppress the SM backgrounds. In such analyses, the reconstruction of the top quark becomes complicated as the information on the neutrino originating from top decay is not directly available. It has to be inferred from the total missing energy of the system using different strategies. Additionally, the leptonic decay of the top quark has a smaller branching ratio, which reduces the overall signal yield. These limitations can be mitigated by focusing on the hadronic decay of the top quark, which has a larger branching ratio and where all final-state particles are visible, facilitating complete kinematic reconstruction at the LHC. Furthermore, in scenarios where the top quark is highly boosted, its decay products can be clustered into a single large-radius fat jet. Efficient tagging of such jets not only simplifies reconstruction but also suppresses contributions from the QCD multijet background. These advantages make the study of hadronic top decays particularly compelling for high-energy and high-luminosity hadron colliders. 

Over the past few decades, top tagging has seen remarkable advancements. Traditional top tagging algorithms rely on physics-motivated high-level features (HLFs), the jet substructure observables \cite{Plehn:2011tg, Kaplan:2008ie, Thaler:2008ju, Thaler:2010tr, Thaler:2011gf, Marzani:2019hun, Plehn:2009rk, Plehn:2010st, Larkoski:2014wba, Butterworth:2008iy, Salam:2010nqg, Dasgupta:2015yua}. More recently, with the introduction of machine learning (ML) techniques, this field has advanced even further. ML-based classifiers have leveraged HLFs to construct more discriminating features through algorithms like boosted decision trees \cite{Bhattacherjee:2022gjq, Bhattacharya:2020aid}, and multi-layer perceptrons (MLPs) \cite{Chakraborty:2020yfc}. Beyond HLFs, modern ML techniques have expanded to utilize low-level features (LLFs), which include raw or minimally processed jet constituent information. Classifiers trained on LLFs, such as convolutional neural networks (CNNs) \cite{Kasieczka:2017nvn, Macaluso:2018tck, Choi:2018dag, CMS-DP-2017-049}, recurrent neural networks (RNNs) \cite{Egan:2017ojy}, recursive neural networks (RvNNs) \cite{Dreyer:2020brq}, and graph neural networks (GNNs) \cite{Gong:2022lye, Bogatskiy:2020tje, Qu:2019gqs, Moreno:2019bmu, Mikuni:2021pou, Konar:2021zdg, Shimmin:2021pkm} have shown superior performance compared to traditional HLF-based classifiers as they can directly exploit the fine-grained granularity of LHC tracking detectors and calorimeters which allows them to capture more detailed information. Furthermore, recent analyses \cite{Sahu:2023uwb} demonstrate that combining HLFs and LLFs using ensemble models can achieve even better performance with lower uncertainty for different Monte-Carlo event generators, showcasing the potential of hybrid strategies in top tagging.

In this analysis, we employ LorentzNet \cite{Gong:2022lye}, a Lorentz and permutation-equivariant GNN, as our top quark tagger. LorentzNet takes the four-momentum of fat jet constituents along with additional Lorentz-invariant features, such as the constituent masses, as inputs. The original LorentzNet model was trained on datasets of top and QCD fat jets with transverse momentum ($p_T$) in the range of 550 to 650 GeV. While this pre-trained model is publicly available, our study requires a tagger capable of distinguishing top and QCD fat jets over a broader $p_T$ range, as relevant to our analysis. Additionally, the dataset used to train the original LorentzNet does not incorporate tracking information from the LHC detectors, which has been shown to significantly enhance tagging performance \cite{Sahu:2023uwb}. To address these limitations, we regenerate training and evaluation datasets following the methodology of Ref. \cite{Sahu:2023uwb}, extending the $p_T$ range and incorporating tracking information. This refined dataset ensures the adaptability of LorentzNet to the needs of our analysis.

The article is organised as follows. In section~\ref{sec:model} we briefly discuss the simplified model we consider for our study. In section~\ref{sec:collider_analysis} we discuss our strategy for collider analysis in detail followed by section~\ref{sec:result} where we discuss our results and the impact it has on SUSY searches. Finally, in section~\ref{sec:concl}, we summarise our work and conclude.

\section{Model definition}
\label{sec:model}
Introducing R-parity violation (RPV) in the theory leads to either lepton number or baryon number violation by one unit. Lepton number and baryon number violation together can lead to proton decay and put stringent constraints on the 
RPV couplings~\cite{Dreiner:1997uz}. The generic superpotential in the presence of R-parity violation can be extended to include the following terms \cite{Dreiner:1997uz,Barbier:2004ez,Banks:1995by}

\begin{equation} 
	\label{eq:rpv_potential}
	W_{\slashed{R}_p} = \mu_iH_u.L_i + \frac{1}{2}\lambda_{ijk}L_i.L_je_k^c + 
	\lambda^\prime_{ijk}L_i.Q_jd_k^c + \frac{1}{2}\lambda^{\prime\prime}_{ijk}u_i^cd_j^cd_k^c
\end{equation}

Here L (e), Q, u (d), and $H_u$ represent left (right) handed lepton superfield, left handed quark doublet, right-handed singlet up-type (down-type) quark superfield, and up-type Higgs superfield respectively.  Generation indices are denoted by i, j, k and $c$ denotes charge conjugation. Lepton number is violated by the first three terms in Eq.~\ref{eq:rpv_potential} while baryon number is violated by the last term. In this analysis, we work with a simplified scenario where all the RPV terms except $\lambda^{\prime\prime}_{323}$ are set to zero. 

\begin{figure}[h!]
	\centering
	\includegraphics[width=0.48\columnwidth]{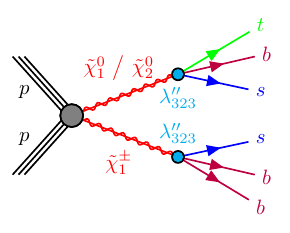}
	\includegraphics[width=0.48\columnwidth]{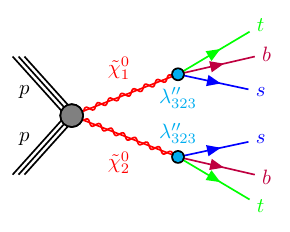}
	\caption{Production and decay modes of the pure higgsino states in the simplified model considered in this work.}
	
	\label{fig:prod}
\end{figure}

Higgsino serves as the LSP in this scenario. All three higgsino states, being pure in nature, are mass degenerate. The bino and wino parameters are thus set to be completely decoupled from the higgsino sector. In this simplified model, both the neutral higgsinos ($\tilde{\chi}_1^0$ and $\tilde{\chi}_2^0$) have a common decay mode, that consists of a top, a bottom and a strange quark. On the other hand the charged higgsino ($\tilde{\chi}_1^{\pm}$) decays into a strange and two bottom quarks. The higgsino production and decay modes are shown in Fig.~\ref{fig:prod}.
\section{Collider Analysis}
\label{sec:collider_analysis}

In this section, we present the phenomenological aspects of Higgsino production 
in the RPV SUSY scenario with UDD type couplings.  As outlined previously, our model features mass degenerate $\lspone, \lsptwo$ and $\chonepm$, while 
all other SUSY particles are decoupled with masses set at 10 TeV for our analysis. 
Consequently, at the LHC, only the production of these electroweakinos is expected to yield a measurable cross-section. Our study will focus on three key production channels for these Higgsinos: $pp\rightarrow \tilde{\chi}_1^0 \tilde{\chi}_2^0$, $pp\rightarrow \tilde{\chi}_1^0 \tilde{\chi}_1^{\pm}$, and $pp\rightarrow \tilde{\chi}_2^0 \tilde{\chi}_1^{\pm}$.
Based on the ATLAS collaboration's analysis of RUN-II LHC data for non-zero 
$\lambda^{\prime\prime}_{UDD}$ couplings, neutralino masses up to 320 GeV have already been excluded \cite{ATLAS:2021fbt}. Therefore, our focus will be on the mass range between 400 and 1000 GeV.

Given the masses of these electroweakinos, their decays may result in the production of one or more boosted top quarks at the LHC. In this work, we propose a search strategy that leverages these boosted final state objects to enhance the discovery potential. Our analysis is centered on assessing the discovery prospects of the RPV SUSY scenario at the future 14 TeV High-Luminosity LHC (HL-LHC).



\subsection{Event Simulation and Object Reconstruction}
\label{sec:obj_recon}
To generate signal events, our analysis utilizes the default RPV-MSSM model file of \texttt{SARAH- 4.14.5} \cite{SARAH1,SARAH2} and uses \texttt{SPheno-4.0.4} \cite{SPheno1,SPheno2} for the generation of SUSY particle spectrum. As the background for the present analysis, we consider relevant SM processes such as diboson, triboson, tetraboson production, and multi-top productions in association with/without gauge bosons. The generation of electroweakino pairs and the background processes, accompanied by up to two additional partons, are performed at the leading order using \texttt{MadGraph5-aMC@NLO} \cite{MadGraph}. Subsequent decays of unstable particles, parton showering, fragmentation, and hadronization processes are simulated with \texttt{PYTHIA 8.2} \cite{Pythia8.2}. Our analysis accounts for both initial and final state radiation, while multi-parton interactions and pile-up effects are neglected for simplicity. Detector effects are modeled using the fast detector simulation tool \texttt{Delphes-3.4.2} \cite{Delphes}, utilizing the default ATLAS detector configuration.\\

In the final analysis, both small-radius and large-radius jets are employed. Jet reconstruction is performed using \texttt{FastJet} \cite{FastJet} with the \textit{anti-$k_T$} algorithm \cite{Anti-KT}, utilizing a radius parameter of $R=0.4$ for small jets and $R=1.2$ for large, or ``fat" jets. The small-radius jets are required to have a transverse momentum ($P_T$) greater than 25 GeV and a pseudorapidity ($\eta$) less than 2.8. For fat jets, we consider only those with $P_T > 300$ GeV and $\eta < 2.8$. Additionally, a top-tagging requirement, based on a Graph Neural Network (GNN) top tagger, is applied to these fat jets (details are provided in Appendix \ref{sec:tag_model}). In the final analysis, only fat jets tagged as top jets are retained, while other fat jets are excluded. To prevent double counting, any $R=0.4$ jets within a distance of $R=1.2$ from a top-tagged fat jet are removed.\\

Electron candidates are selected with a transverse momentum threshold of $P_T > 25$ GeV and a pseudorapidity range of $|\eta| < 2.47$. Additionally, they must meet a `loose' isolation criterion as outlined in Ref. \cite{ATLAS:2019jvq} and are excluded if located in the transition region between the electromagnetic calorimeter (ECAL) barrel and endcap. Muon candidates are required to have a $P_T > 25$ GeV, a pseudorapidity range of $|\eta| < 2.7$, and must also satisfy a `loose' isolation criterion as defined by the ATLAS collaboration \cite{ATLAS:2020auj}. The missing transverse momentum  ($\met$) is calculated as the negative vector sum of the transverse momenta of all reconstructed objects and the tracks that are not associated with these objects. To prevent potential double-counting of selected objects, a dedicated procedure based on prescription by the ATLAS collaboration \cite{ATLAS:2016poa, ATLAS:2020uiq} is employed.\\

\subsection{Event Selection}
\label{event_selection}
To separate the signal and background events effectively, for our final analysis, we devise two signal regions (SR):
\begin{itemize}
	\item \texttt{SR1:} ($N_{top} \geq 1$) $\cap$ ($N_{bjet} \geq 3$) $\cap$ ($N_{light-jet} \geq 2$) $\cap$ ($N_{lep}=0$)
	\item \texttt{SR2:} ($N_{top} \geq 1$) $\cap$ ($N_{bjet} \geq 1$) $\cap$ ($N_{light-jet} \geq 2$) $\cap$ ($N_{lep}=1$)
\end{itemize}
Here, $N_{top}$, $N_{bjet}$, $N_{light-jet}$, and $N_{lep}$ represent the number of top-tagged jets, b-tagged jets, light jets, and leptons in the signal/background events, respectively. These two signal regions are mutually independent. Therefore, the median expected significance in these SRs can be combined statistically.
The \texttt{SR1} is designed to capture event topologies where all the electroweakinos decay hadronically. Since all final state particles are visible, it is possible to reconstruct the masses of the parent Higgsinos by carefully combining the final state reconstructed objects, i.e., jets. Additionally, this reconstructed invariant mass can serve as a very important discriminating variable between the signal and background. The \texttt{SR2} primarily focuses on the production channel $pp\rightarrow \chi_1^0 \chi_2^0$, where at least one of the tops coming from the neutralinos decays leptonically. Together, these two signal regions allow us to capture all possible final states with a boosted top quark. 

Based on the characteristics of signal and background events in each signal region, we construct several kinematic variables that can effectively differentiate between the signal and background events. Later, we use these variables to train a boosted decision tree (BDT) to separate signal events from the background ones. In the following, we will discuss a few of the important kinematic variables while leaving the rest to Appendix \ref{kinematic_variables}.

Three of the most effective variables for \texttt{SR1} are the effective mass ($M_{eff}$) of the system, the reconstructed mass of the Higgsinos ($\lspone/ \lsptwo / \chonepm$), and pseudo-top mass ($M_{t^\prime}$). We define the effective mass as the scalar sum of the transverse momentum of all the visible final state particles and the missing transverse momentum ($M_{eff}=\sum P_T^{\rm jets} + \sum P_T^{\rm lepton} + {\met}$). Considering the higher masses of the  Higgsinos involved, we expect $M_{eff}$ to have a greater value for the signal events compared to the background. 

To reconstruct the invariant mass of the SUSY particles, we introduce a novel combinatorial approach. First, we consider selected combinations of 3 jets from the 6 SR1 jets, which include a fat top jet ($t$), two light quark jets ($j_1$ and $j_2$), and three $b$-quark jets ($b_1$, $b_2$, and $b_3$). Since there are 6 jets defining SR1, once a particular combination of 3 jets is grouped together, a second group of 3 jets emerges without any ambiguity. Out of the 120 possible combinations, a further requirement on the group of 3 jets containing the top-tagged jet, that it should include exactly one $b$-tagged jet and one light jet, reduces the number of allowed combinations to 6. The rationale for considering only combinations such as $(t b_1 j_1, b_2 b_3 j_2)$, $(t b_1 j_2, b_2 b_3 j_1)$, and so on lies in the fact that SR1 is designed to probe the final state topologies with a neutralino decaying into $tbs$. From these 6 possible combinations, we select the particular combination for which the invariant masses of the two groups of 3 jets are closest to each other. The rationale is that since the two SUSY particles produced are degenerate in mass, the group with the closest matching invariant masses is likely to contain the jets originating from their decays. This approach helps isolate the jets originating from the SUSY particles and accurately reconstruct their invariant mass.

\begin{figure}[htb!]
		\centering	
		\includegraphics[width=0.32\columnwidth]{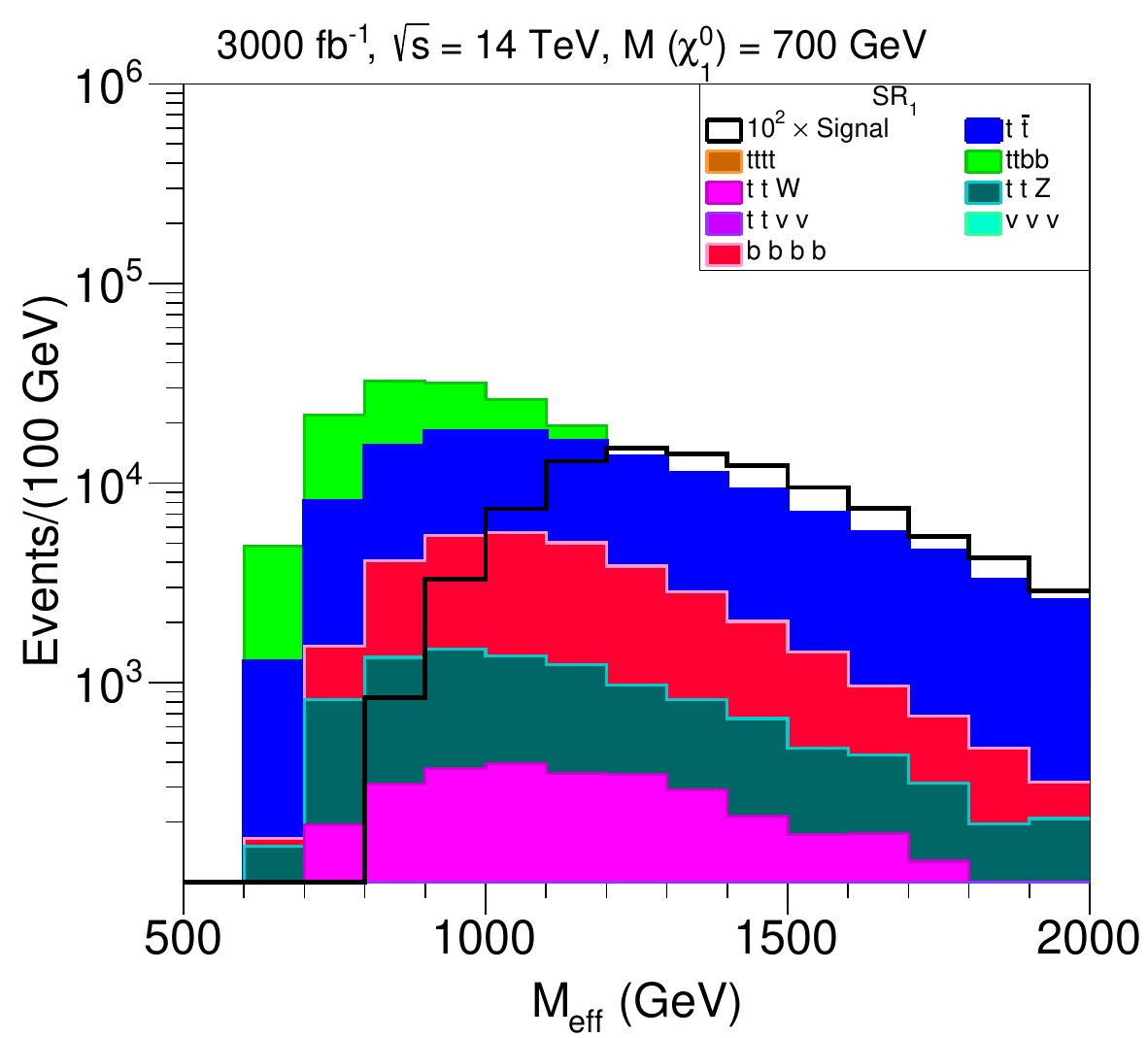}
		\includegraphics[width=0.32\columnwidth]{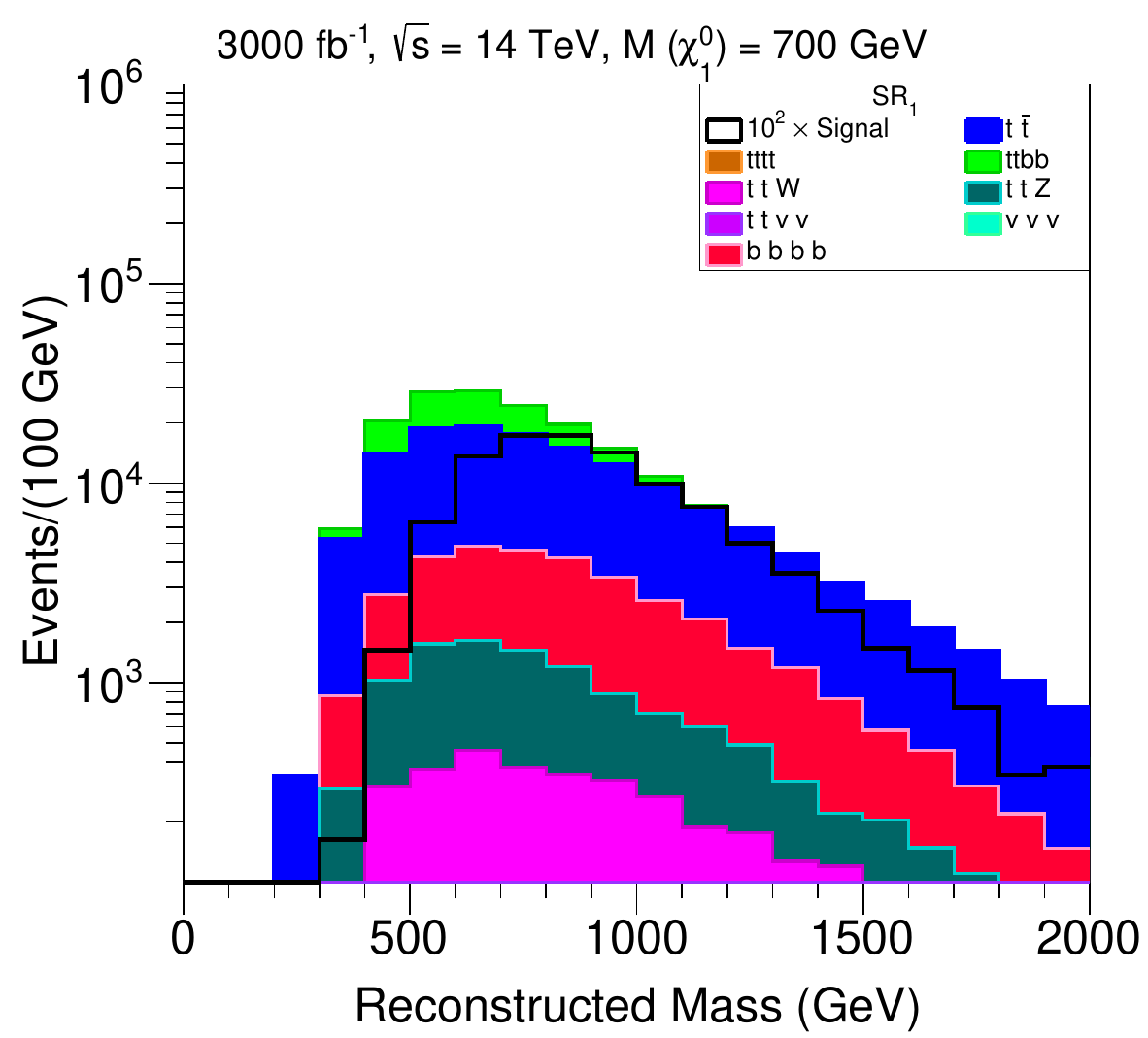}
		\includegraphics[width=0.32\columnwidth]{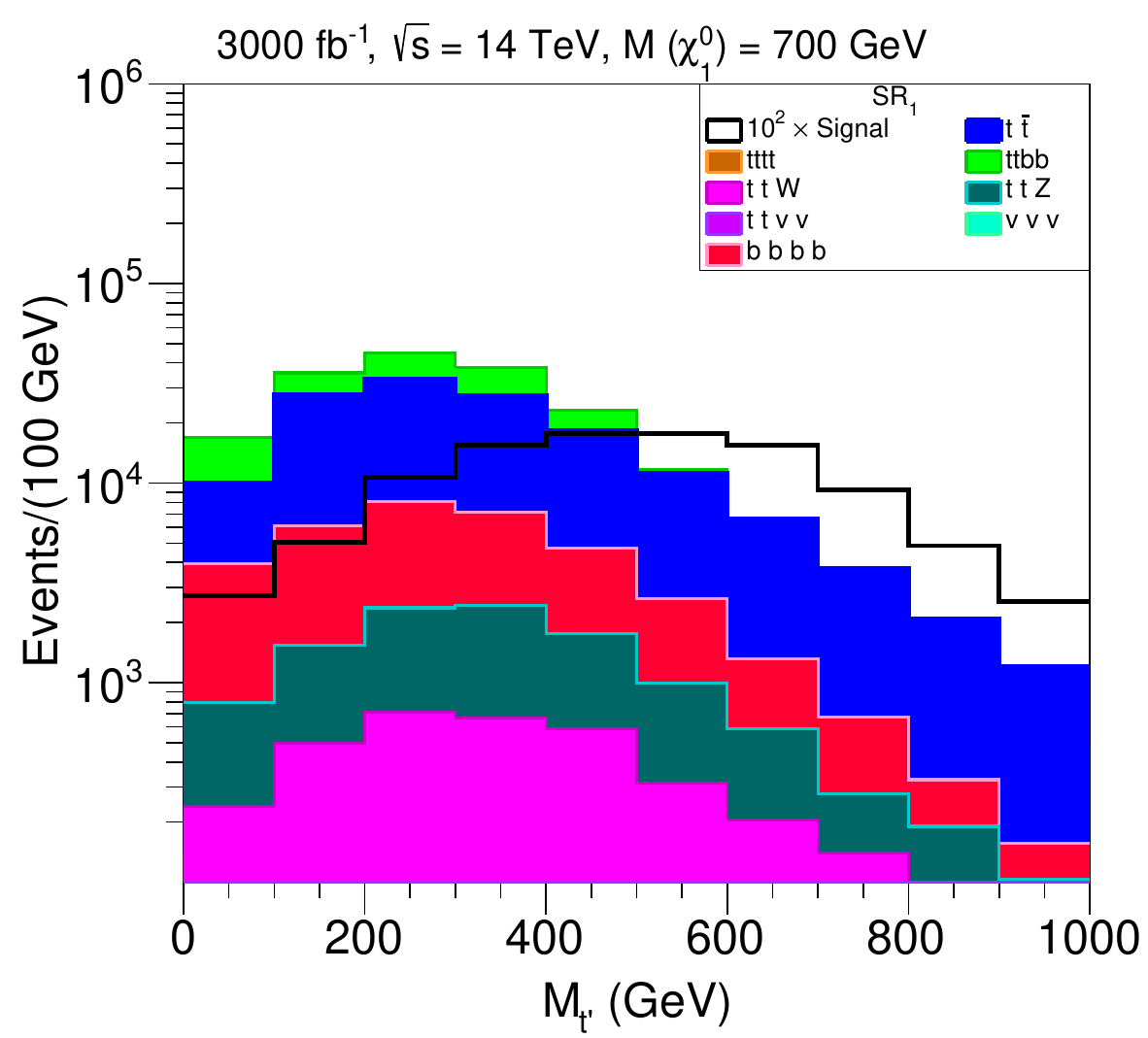}
		\caption{Distributions for the combined signal with $\mlspone= \mchonepm =\mlsptwo=700$ GeV and background events. From left to right, we have the $M_{eff}$, Reconstructed Mass and  $M_{t^\prime}$ for \texttt{SR1}. }
		\label{fig: dist_1}
	\end{figure}
\begin{figure}[htb!]
		\centering
		\includegraphics[width=0.32\columnwidth]{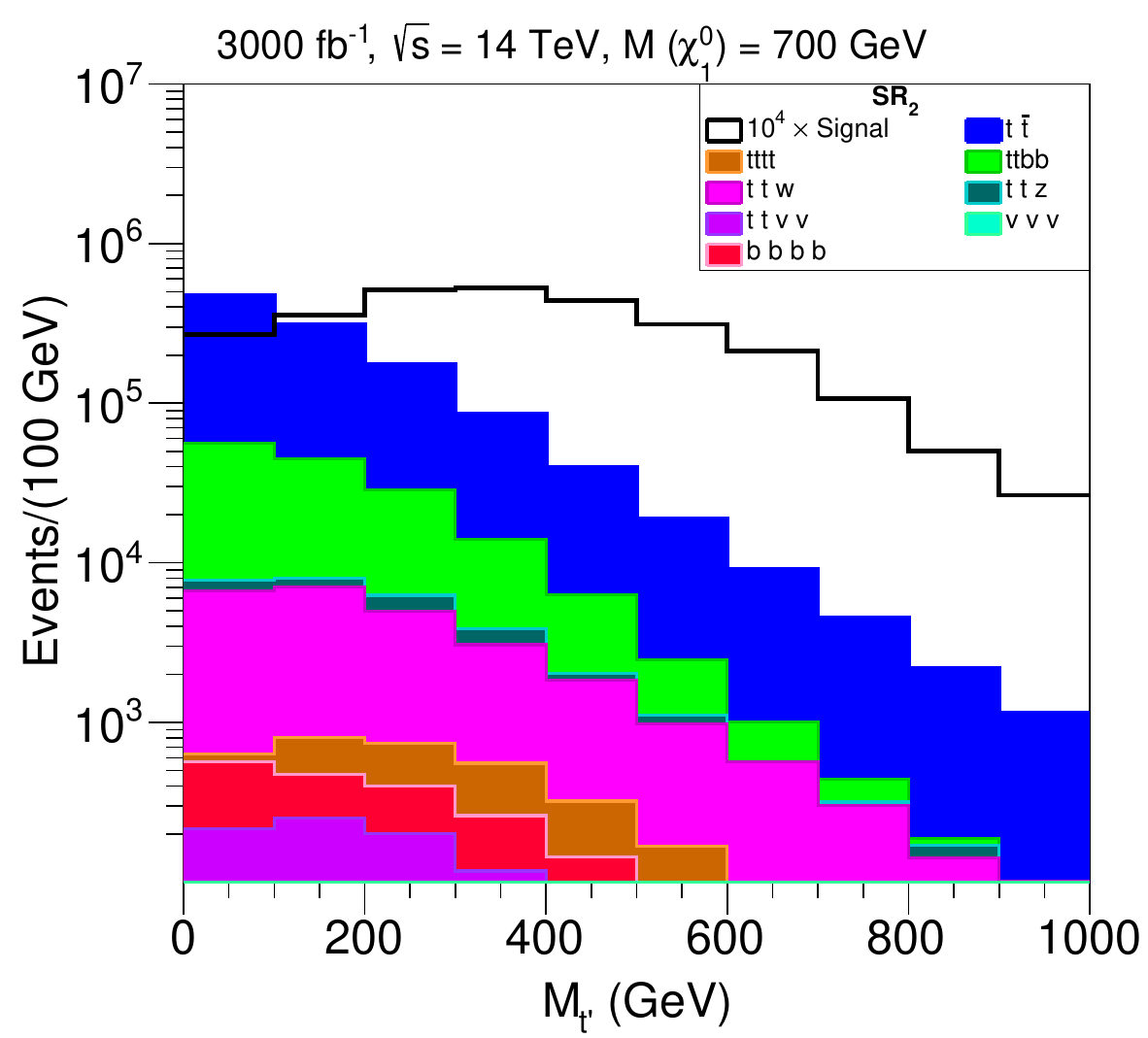}
		\includegraphics[width=0.32\columnwidth]{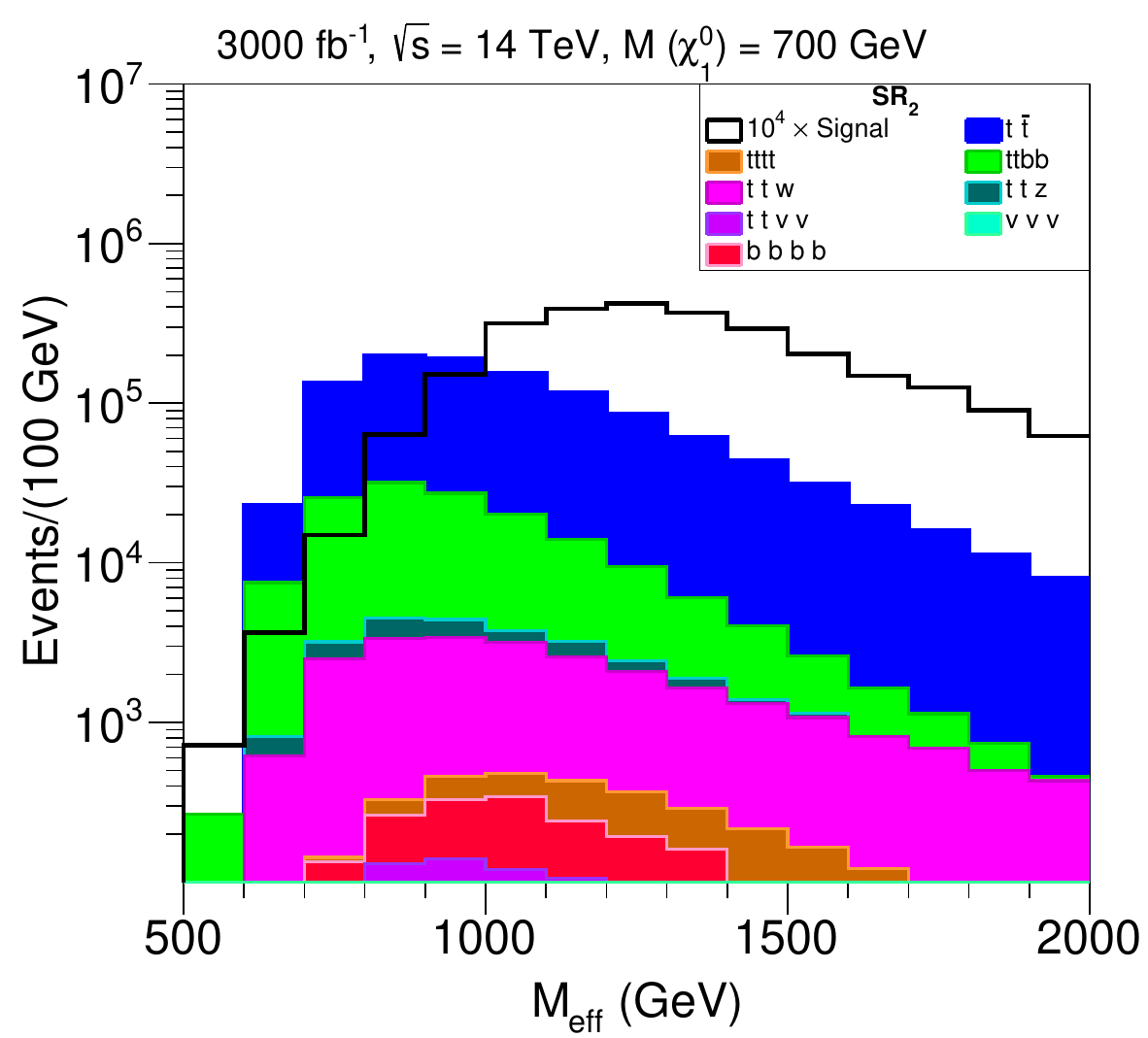}
		\includegraphics[width=0.32\columnwidth]{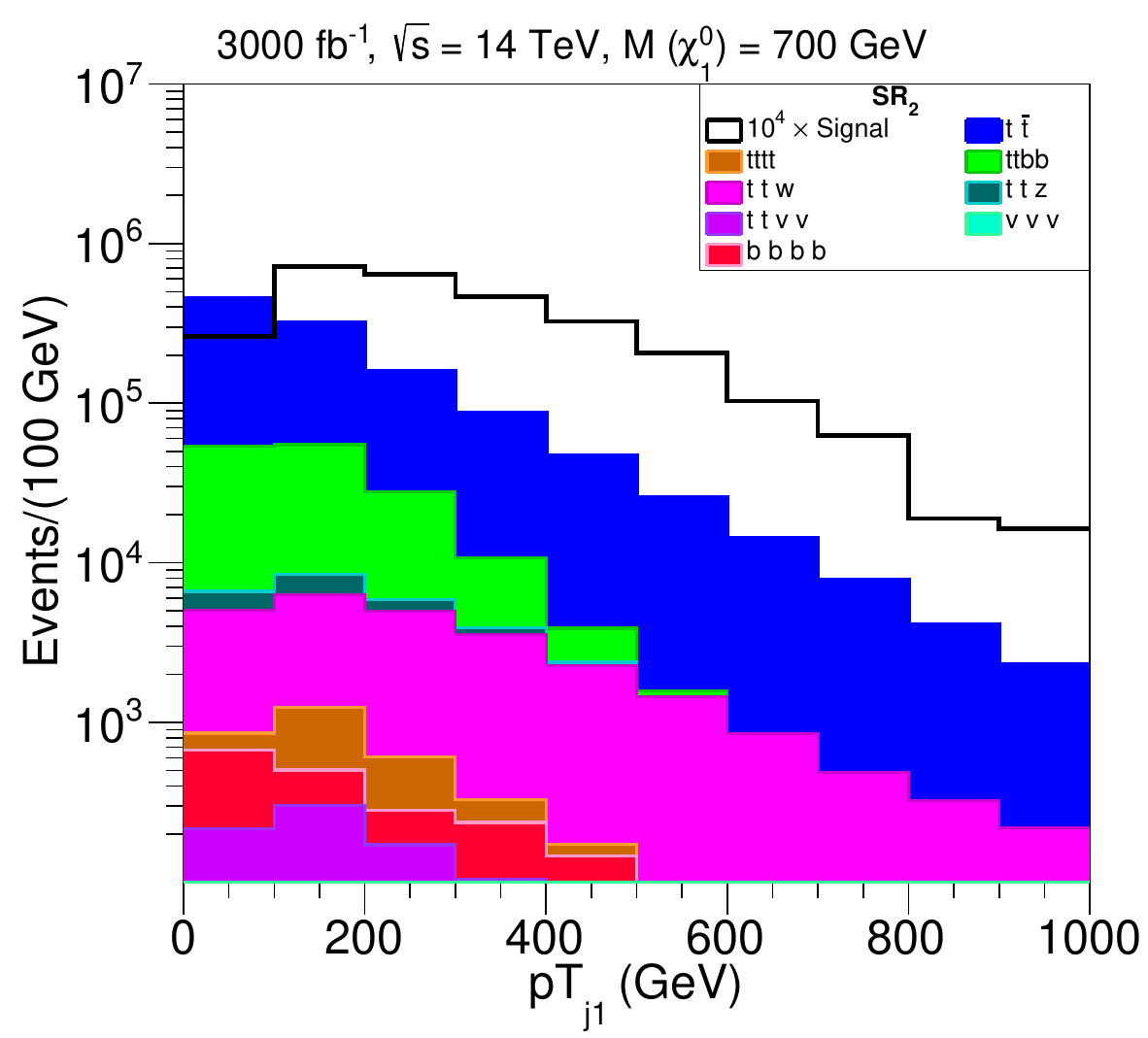}
		\caption{Distributions for the combined signal with $\mlspone= \mchonepm =\mlsptwo=700$ GeV and background events. From left to right, we have the  $M_{t^\prime}$, $M_{eff}$ and $P_T^{j_1}$ for \texttt{SR2}. }
		\label{fig: dist_2}
	\end{figure}

We define the pseudo-top mass $M_{t^\prime}$ as the invariant mass of the jets lying inside a cone of radius 2.0 opposite to the top jet. This variable effectively reduces backgrounds (like top-pair production), where we expect the two tops to be approximately opposite to each other. We present the rest of the variables used for training the BDT in Appendix \ref{kinematic_variables}. We present the distribution of $M_{eff}$, the reconstructed mass of the Higgsinos, and $M_{t^\prime}$ for the \texttt{SR1} signal region in Figure \ref{fig: dist_1} for a Higgsino mass of 700 GeV. For convenience, we have merged the kinematical distributions corresponding to the final states arising from three different signal processes into one. Since all the Higgsinos are mass degenerate, the distributions are expected to be similar. Note that, since the number of signal events per bin is much smaller compared to the SM background channels, the signal event numbers in each bin are scaled by a factor of $10^2$ in Fig.~\ref{fig: dist_1} and $10^4$ in Fig.~\ref{fig: dist_2}. As evident, the signal and background distributions look quite similar and hence a traditional cut-based method is unable to enhance the signal to background ratio significantly. We therefore adapted a machine learning model in order to differentiate between signal and background events. \\

Similarly, the three most effective kinematic variables for the \texttt{SR2} are the transverse momentum of the leading light jet, effective mass $M_{eff}$, and the pseudo-top mass $M_{t^\prime}$. The definition of the $M_{eff}$ and $M_{t^\prime}$ variables are the same as the one used for the \texttt{SR1} . Figure \ref{fig: dist_2} shows the distribution of these variables for the signal (corresponding to 700 GeV electroweakinos) and background samples. The details of the remaining kinematic variables are in Appendix \ref{kinematic_variables}.  
We also show the feature importance plot for the top ten most important 
variables for both the signal regions in Figure \ref{fig:var_imp} 
in Appendix \ref{kinematic_variables}. 
For our final analysis, we select  twelve benchmark points with electroweakino masses ranging from 400 GeV to  950 GeV in 50 GeV increments. 
The signal cross-sections at the NLO+NLL level are calculated using \texttt{Resummino-3.1.1} \cite{Resummino}. For the Standard Model background processes, we apply NLO K-factors \cite{Muselli:2015kba, Catani:2009sm, Balossini:2009sa, Broggio:2019ewu, Kidonakis:2015nna, Campbell:2011bn, Shen:2015cwj} to scale the LO cross-sections, ensuring an accurate representation of the background contributions.

\section{Results}
\label{sec:result}
In this section, we will discuss the future projection for the RPV-MSSM scenario considered in our analysis. Our results correspond to the high luminosity (3000 $fb^{-1}$) and high energy ($\sqrt{s}=14$ TeV) run of the LHC. 
As outlined in the previous section, we have defined two distinct signal regions to effectively manage Standard Model background contributions and capture all signal events involving at least one final-state boosted top jet. The key kinematic variables for distinguishing signal from background are also discussed earlier and in Appendix \ref{kinematic_variables}. These variables have been used in our final analysis to train BDT-based classifiers. For each neutralino mass value between 400 GeV and 1000 GeV, at 50 GeV intervals, two classifiers are trained corresponding to the two signal regions. All classifiers share the same set of hyperparameters and are implemented using the {\tt TMVA~4.3} toolkit \cite{Hocker:2007ht} of {\tt ROOT~6.24} \cite{Brun:2000es}. Further details on these models are provided in Appendix \ref{hyper_params}.

The preselection criteria for the two signal regions are very strict and reduce the number of background events in the sample drastically (although the final yield still remains significantly large compared to the signal yield owing to the large background cross-section). However, a sufficient number of background events is crucial for properly training the classifiers. For our analysis, we have generated enough background samples to ensure that 100,000 events remain after preselection. These events are weighted according to their cross-sections and used for classifier training. As for the signal events, note that three different production channels can contribute to the final states considered here.  We have similarly generated enough signal events so that, after preselection, 100,000 signal events remain for training. Relative weights were also introduced among the signal channels to reflect their contributions. In the following, we present the results of our BDT-based classification. For brevity, we focus on the classifiers trained on the signal corresponding to a neutralino mass of 700 GeV.

\begin{figure}[h!]
	\centering
	\includegraphics[width=0.49\columnwidth]{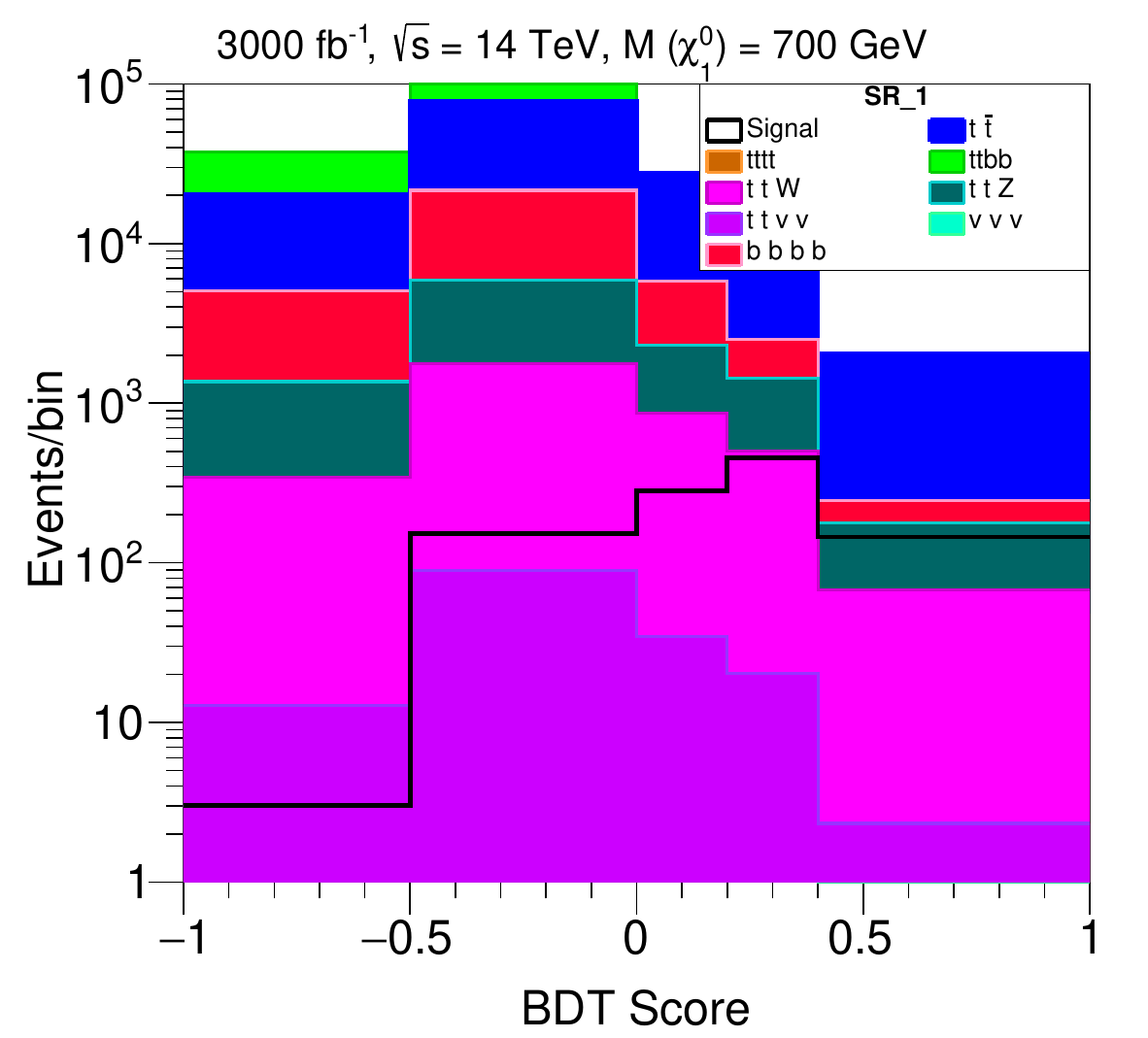}
	\includegraphics[width=0.49\columnwidth]{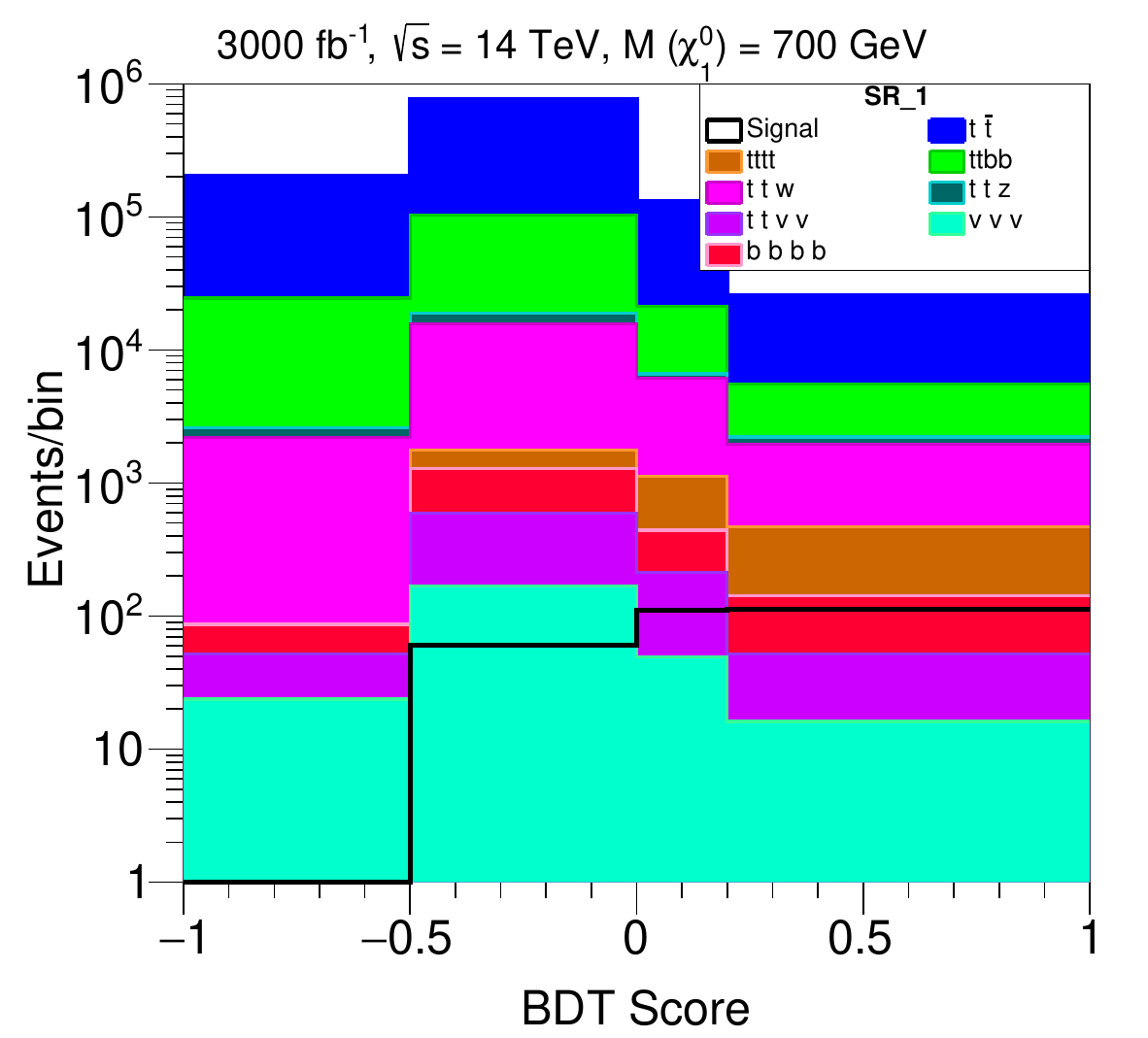}
	\caption{Distributions for the combined signal events with $\mlspone=700$ GeV and background events as a function of BDT score for the signal regions \texttt{SR1}(left) and \texttt{SR2}(right).} 
	\label{fig: bdt_score}
\end{figure}

In Figure \ref{fig: bdt_score}, we present the distribution of the scores of two BDT classifiers for \texttt{SR1} (left) and \texttt{SR2} (right). As mentioned above, these plots correspond to classifiers trained with signal events generated with a Higgsino mass of 700 GeV. For completeness, we have presented the distribution of each background separately. In the signal histogram, we have combined the contribution from the three production channels. 
In our analysis, these BDT scores serve as the final discriminating observable. For each value of neutralino mass, we divide the distribution of the BDT score for \texttt{SR1}  into five bins with edges at $-1, -0.5, 0, 0.2, 0.4, 1$ and the BDT score of \texttt{SR2}  into four bins with edges at $-1, -0.5, 0, 0.2, 1$. We calculated the median expected exclusion significance in each of the nine bins and added them in quadrature to obtain the final significance. \\

\begin{figure}[!htb!]
	\centering
	\includegraphics[width=0.85\columnwidth]{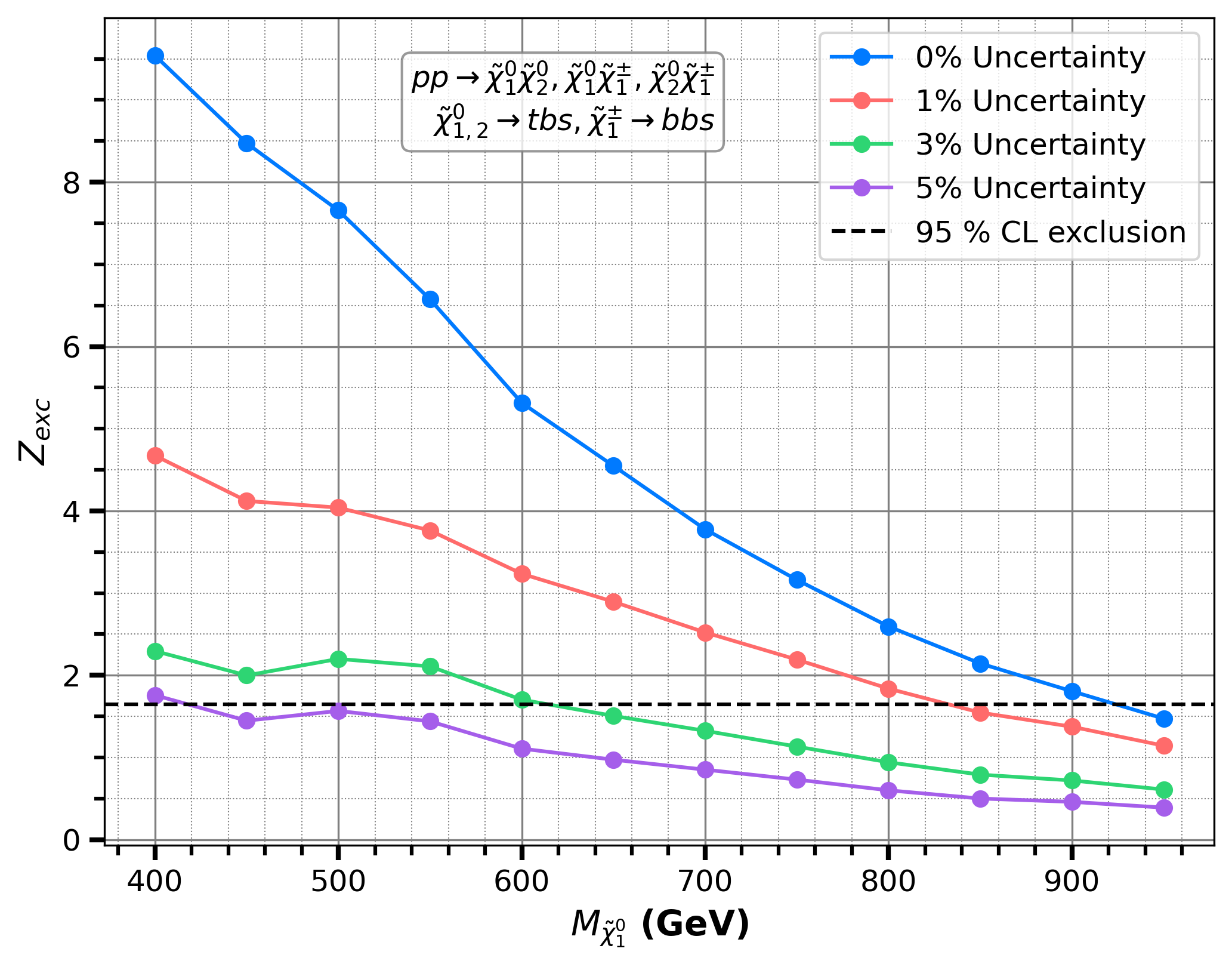}
	\caption{Median expected exclusion significance for the RPV frameworks with Higgsino pair production at the future HL-LHC ($\sqrt{s}=14$ TeV, $\cal{L}$ = 3000 $fb^{-1}$). Here the classifiers are tested and trained on the same mass of Higgsino for the signal events for each benchmark point. 
		The blue, red, green, and violet colors correspond to expected exclusion contours with 0\%, 1\%, 3\%, and 5\% background uncertainties. 
	}
	\label{fig:res_1}
\end{figure}

Following the Refs. \cite{Cowan:2010js,Li:1983fv,Cousins:2007yta}, we used the following expression to calculate the median expected exclusion significance:
\begin{equation*}
    Z_{\rm exc} = \left[ 2 \left\{ s-b \ln \left( \frac{b+s+x}{2b} \right) - \frac{b^2}{\delta_b^2} \ln \left( \frac{b-s+x}{2b} \right) \right\} - (b+s-x)(1+b/\delta_b^2) \right]^{1/2}
\label{eq:zdis2}
\end{equation*}
Here, $x = \sqrt{(s+b)^2 - 4sb\delta_b^2/(b+\delta_b^2)}$, $s$ and $b$ are the numbers of signal and background events, respectively, and $\delta_b$ is the uncertainty in the measurement of the background. The exact estimation of background uncertainty is beyond the scope of our analysis. We have adopted a conservative approach and present our result for three different values of this uncertainty: 1\%, 3\%, and 5\%.

\begin{figure}[!htb]
	\centering
	\includegraphics[width=0.85\columnwidth]{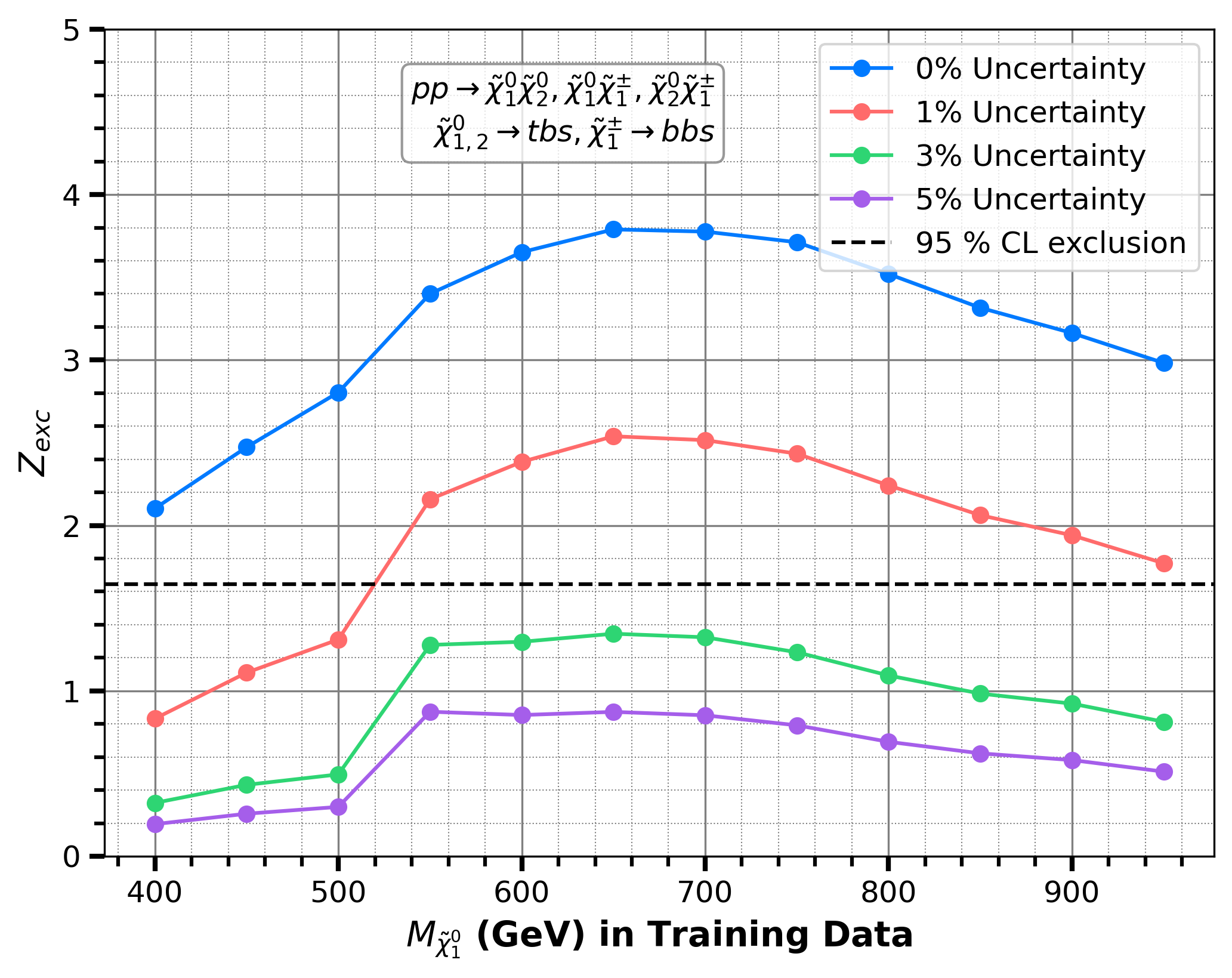}
	\caption{Median expected exclusion significance for the RPV frameworks with Higgsino pair production at the future HL-LHC ($\sqrt{s}=14$ TeV, $\cal{L}$ = 3000 $fb^{-1}$). Here the classifiers are trained with the mass of neutralino corresponding to the benchmark point, but tested on the 700 GeV signal sample. The color conventions are similar to Figure \ref{fig:res_1}. }
	\label{fig:res_2}
\end{figure}

In Figure \ref{fig:res_1}, we present the median expected exclusion significance for different values of Higgsino mass. Note that to obtain these results, we tested the classifiers trained with a given Higgsino mass with signal events of the same mass. The results for 0\%, 1\%, 3\%, and 5\% background uncertainties are presented in blue,  red, green, and violet, respectively.
Our results demonstrate that  without any background uncertainties, it is possible to probe the RPV SUSY model up to around 925 GeV Higgsino mass at a 95\% C.L., which corresponds to a $Z_{esc} = 1.645$. As we keep on increasing the uncertainty, the reach diminishes, and for uncertainties above 5\%, it is only possible to exclude the 400 GeV neutralinos at 95\% C.L.
The signal yield and background yield, along with the signal significance for the different values of background uncertainties for two benchmark points, with Higgsino mass at 500 and 700 GeV, are presented  in Table \ref{data_table}.
in Appendix \ref{yield}.

In Figure \ref{fig:res_2}, we present a similar result as Figure \ref{fig:res_1} with the only difference that here, all the classifiers are tested for the signal sample corresponding to 700 GeV neutralinos. The lines in blue, red, green, and violet correspond to background uncertainties of 0\%, 1\%, 3\%, and 5\%, respectively. As expected, the expected significance is higher for classifiers where the training data has a similar mass to the testing data. Such a setup can be helpful in the future HL-LHC to get a rough estimate of the mass of the BSM particle.

\section{Conclusion}
\label{sec:concl}
High energy and high luminosity LHC is going to probe new physics parameter space like never before. TeV scale supersymmetry has been probed quite meticulously by the ATLAS and CMS collaborations. However, a large portion of the sub-TeV SUSY parameter region still remains unexplored, mainly owing to small a production cross-section and (or) small cut efficiencies. The $\mu$-parameter is vital to the naturalness of a SUSY theory and it is of utmost importance to probe sub-TeV values of $\mu$ with every possible scenario. Unfortunately, a small production cross-section of higgsinos makes this particularly difficult at a collider experiment. In this work, we attempt to address this issue in the context of RPV SUSY with specific baryon number violating trilinear interactions. The higgsino mass in this scenario has already been explored by the experimental collaborations up to 320 GeV. It is difficult to achieve much improvement on the exclusion limit with a similar analysis even at high luminosity LHC. Hence we adopt a different strategy. We aim to tag one of the top jets in the final state as a fat jet in addition to other light jets and $b$-jets and use that to suppress backgrounds. A novel and highly efficient top-tagger based on a GNN has been used for this purpose. As we increase the higgsino mass, one would expect the decay products of the higgsinos to be more boosted. So, it is a trade-off between lower cross-section and higher top tagging efficiency as the higgsino masses get heavier. We combine three different production channels with the two neutral and one charged higgsino and construct two signal regions. The kinematic distributions of the signal events turn out not to be very distinguishable from those of the SM background events. Hence we adapt a machine learning model to identify the finer distinguishing features in order to achieve better identification of the signal and background events. Two BDT-based classifiers are trained to this end using the TMVA toolkit. We combine the statistical significance obtained from the two signal regions to present our final results. We conclude that using our method, one can effectively probe Higgsino masses in the range of 400 - 925 GeV. 

\section{Acknowledgement}
AC, KG and SM acknowledge ANRF India for providing Core Research Grant no. CRG/2023/008570. RB and SM acknowledge ANRF India for providing Core Research Grant no. CRG/2022/003208.

\appendix

\section{Top Tagging}
\label{sec:tag_model}
To tag boosted fat jets originating from top quarks or QCD-initiated light quarks and gluons, we employed the GNN-based classifier LorentzNet \cite{Gong:2022lye}. The publicly available LorentzNet model was originally trained on fat jet samples with transverse momentum ($p_T$) in the range of 550 to 650 GeV. However, for our analysis, we anticipate a broader $p_T$ distribution for top fat jets. Therefore, we generated our own signal and background datasets to train a more customized version of LorentzNet tailored to our study.

The procedure for generating signal and background events mirrors the approach in Ref. \cite{Gong:2022lye}. We generated top and QCD jets covering a wide transverse momentum range from 300 to 1000 GeV. To ensure uniform coverage across this range, we divided the $p_T$ spectrum into seven bins, each 100 GeV wide. For each bin, we produced 200k signal and 200k background samples for training. Additionally, we generated 50k samples for each $p_T$ bin to test and validate the classifier.

Compared to the original LorentzNet dataset, we have introduced two modifications. First, we followed Ref. \cite{Sahu:2023uwb} and incorporated the information from tracker detectors into the dataset. However, unlike in Ref. \cite{Sahu:2023uwb}, we retained the constituent mass information within the four-momentum of the constituents. Second, instead of using the constituent mass as the node scalar (as in the original LorentzNet paper \cite{Gong:2022lye}), we used the constituent charge. This adjustment was made because the mass information is already embedded within the four-momentum, which we provide as input to LorentzNet.

\begin{figure}
	\centering
	\includegraphics[width=0.5\linewidth]{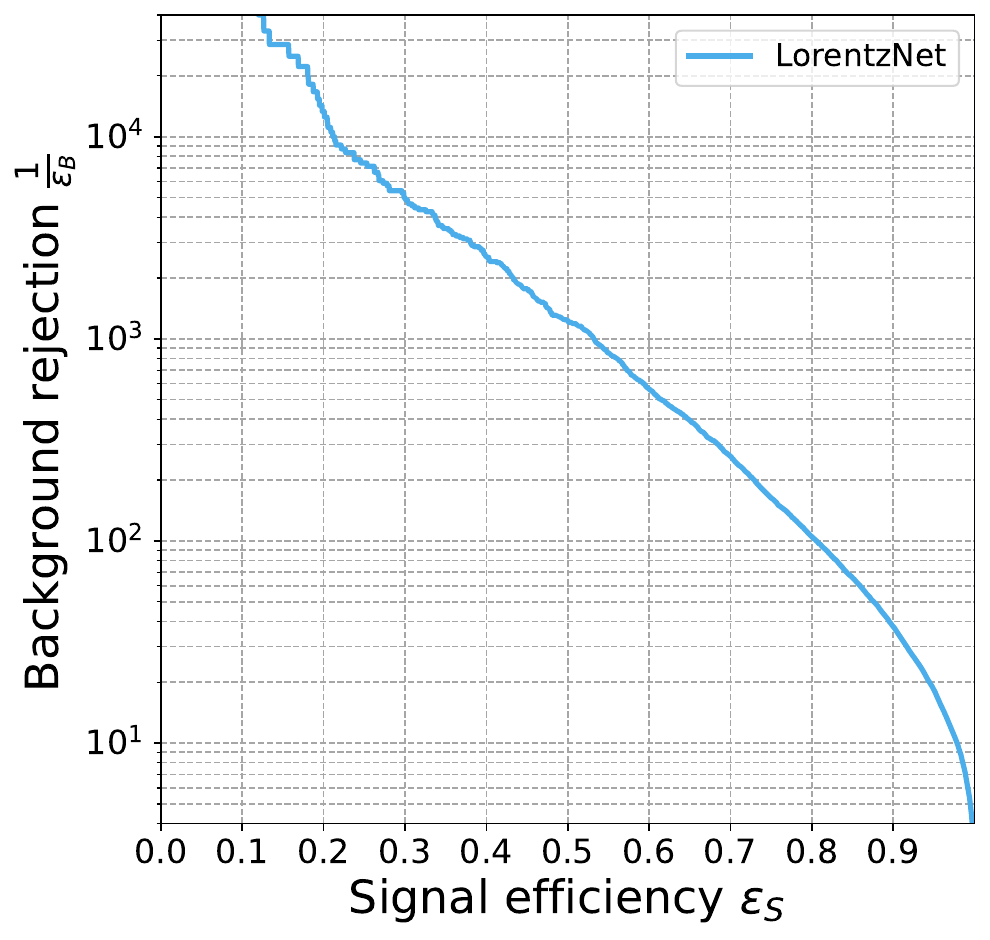}
	\caption{ROC curve of LorentzNet classifier.}
	\label{fig:rocln}
\end{figure}

In Figure \ref{fig:rocln}, we present the Receiver Operating Characteristic (ROC) curve of our LorentzNet classifier, evaluated on top and QCD fat jets with transverse momentum in the range of 500–700 GeV. In our final analysis, we tag $R=1.2$ fat jets as top jets by requiring a GNN score greater than 0.95, which corresponds to a signal efficiency of 56\%. 

\section{Kinematic Variables}
\label{kinematic_variables}
In this section, we will talk about the kinematic variables used as BDT features in the  two signal regions \texttt{SR1} and \texttt{SR2}. A total of 36 variables were used, 23 in \texttt{SR1} and 27 in \texttt{SR2}. The list of all the kinematic variables is shown in Table~\ref{kinematic_variables_table}.

\begin{table}[h!]
	\centering
	\caption{Table of all Kinematic Variables used as features to train the BDTs for each signal region.}
	\label{kinematic_variables_table}
	\renewcommand{\arraystretch}{1}
	
	\begin{tabular}{||c|c|c||} 
		\hline
		\textbf{Kinematic Variables} & \textbf{\texttt{SR1}} & \textbf{\texttt{SR2}} \\ 
		\hline\hline
		
		$P_T^{j_1}$,$P_T^{j_2}$, $P_T^{b_1}$, $P_T^{t_1}$, $\met$, $R(p_T)$ & \multirow{4}{*}{\checkmark} & \multirow{4}{*}{\checkmark} \\
		$M_{eff}$, $H_T$, $M_{t'}$, $\tilde{M}_1^{\tilde{\chi}}$ & & \\
		$N_b$, $N_j$,  $\Delta \phi (j_1, j_2)$, $\Delta \phi (t_1, b_1)$ & & \\
		\hline
		\hline
		
		$M_{CT}$ & \multirow{2}{*}{\checkmark} & \multirow{2}{*}{} \\
		$P_T^{t_2}$, $P_T^{b_2}$, $P_T^{b_3}$ & &\\
		$\Delta \phi (b_1, b_2)$, $\Delta \phi (b_1, b_3)$, $\Delta \phi (b_2, b_3)$, $\Delta \phi (t_1, b_2)$, $\Delta \phi (t_1, b_3)$ & & \\
		\hline
		\hline
		
		$P_T^{lep}$, $\tilde{M}_2^{\tilde{\chi}}$& \multirow{5}{*}{} & \multirow{5}{*}{\checkmark} \\
		$\Delta \phi (b_1, lep)$, $\Delta \phi (b_1, j_1)$, $\Delta \phi (b_1, j_2)$ & & \\
		$\Delta \phi (j_1, lep)$, $\Delta \phi (j_2, lep)$ & & \\
		$\Delta \phi (t_1, j_1)$, $\Delta \phi (t_1, j_2)$, $\Delta \phi (t_1, lep)$ & & \\
		$\Delta  \phi (lep, \met)$, $\Delta \phi (t, \met)$, $\Delta \phi(b_1, \met)$ & & \\
		\hline
	\end{tabular}
\end{table}

\begin{itemize}
	\item $P_T^{X}$ is the transverse momentum of the object $X$.
	\item $\met$ is the missing energy.
	\item $M_{eff}$ is a scalar sum of the transverse momenta of all visible particles (jets, leptons, etc.) and the missing transverse energy $(\met)$.
	\item  $H_T$ is the sum of the transverse momenta of all jets.
	\item $M_{CT}$ is called the contransverse mass. It is defined as:
	$M_{CT} = \sqrt{ \left( E_T^A + E_T^B \right)^2 - \left( \vec{p}_T^A - \vec{p}_T^B \right)^2 }$, where: $E_T^A$ and $E_T^B$ are the transverse energies of objects \( A \) and \( B \); $\vec{p}_T^A$ and $\vec{p}_T^B$ are their transverse momenta.
	\item $N_X$ is the number of $X$ objects in the event.
	\item $\Delta \phi (A,B)$ is the difference in the azimuthal angles (the angle in the transverse plane) between two particles or objects $A$ and $B$.
	\item $R(p_T)$ is defined as the ratio of the scalar sum of $p_T$ of jets in the signal regions to the scalar sum of $p_T$ of all jets in the event.
	\item $\tilde{M}_1^{\tilde{\chi}}$ in \texttt{SR1}is the invariant mass as defined in Section ~\ref{event_selection}. In \texttt{SR2}, we use the same method as in \texttt{SR1}, but instead of six initial pairs, we get only one pair of fat-jet and, lepton and missing energy. Then we make groups of all the jets in the event and calculate the mass, as mentioned in the text. 
	\item $\tilde{M}_2^{\tilde{\chi}}$ in \texttt{SR2} is the invariant mass calculated for the system of the lepton, missing energy, two light jets and one b-jet. 
	
\end{itemize}

$X$, for $P_T^{X}$ and $N_X$ can either be jets or leptons. $j, b, t, lep$ represent light jets, bottom jets, top jets and leptons, respectively. Although the objects $A$ and $B$ in $M_{CT}$ technically can be particles or jets, this variable is generally used between two b jets. In our analysis, we have used this variable in \texttt{SR1} between the two leading b jets. In the case of $\Delta \phi$, $A$ and $B$ can be either jets, leptons or missing energy $\met$.

\begin{figure}[htb!]
	\centering
	\includegraphics[width=0.49\columnwidth]{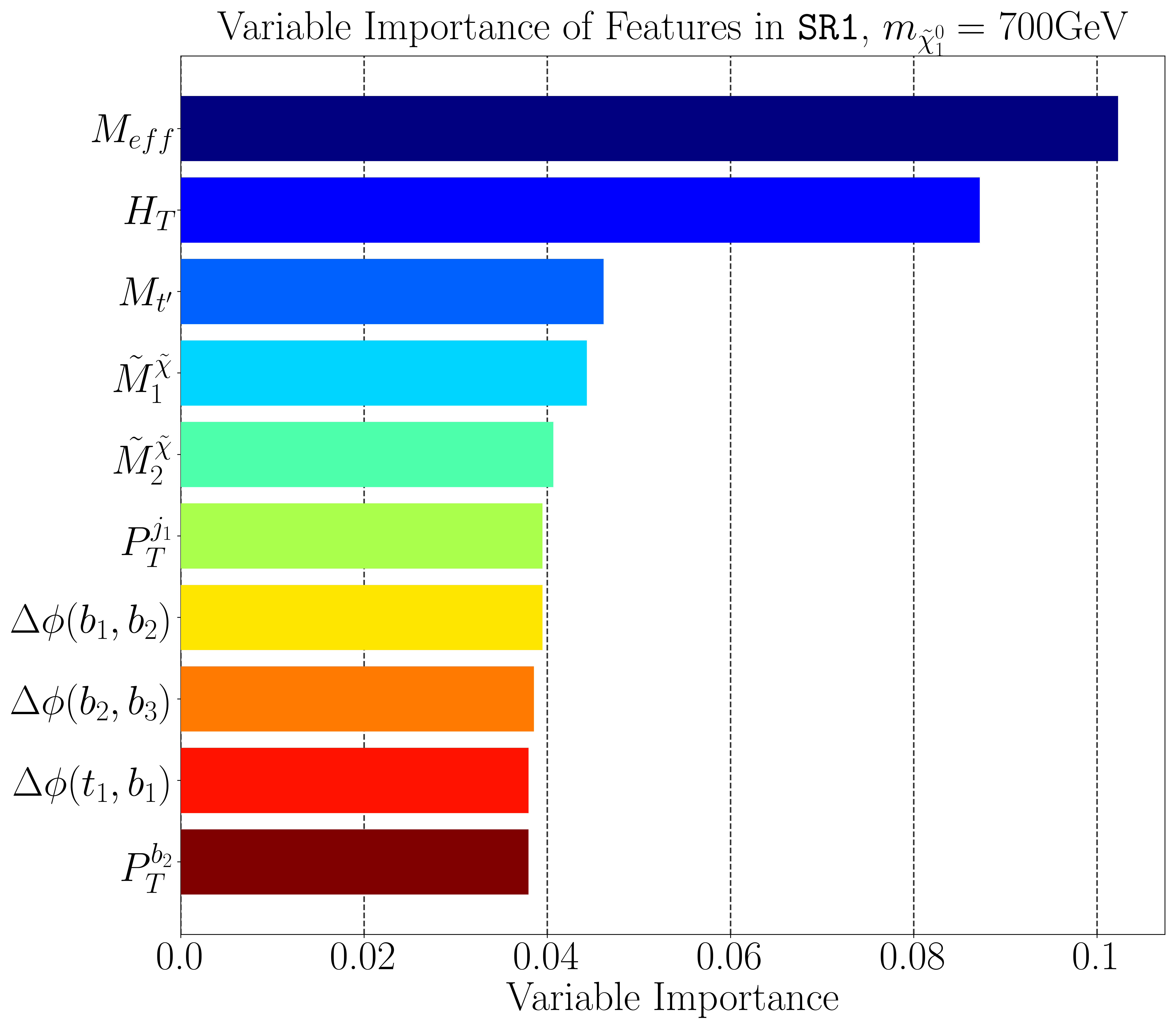}
	\includegraphics[width=0.49\columnwidth]{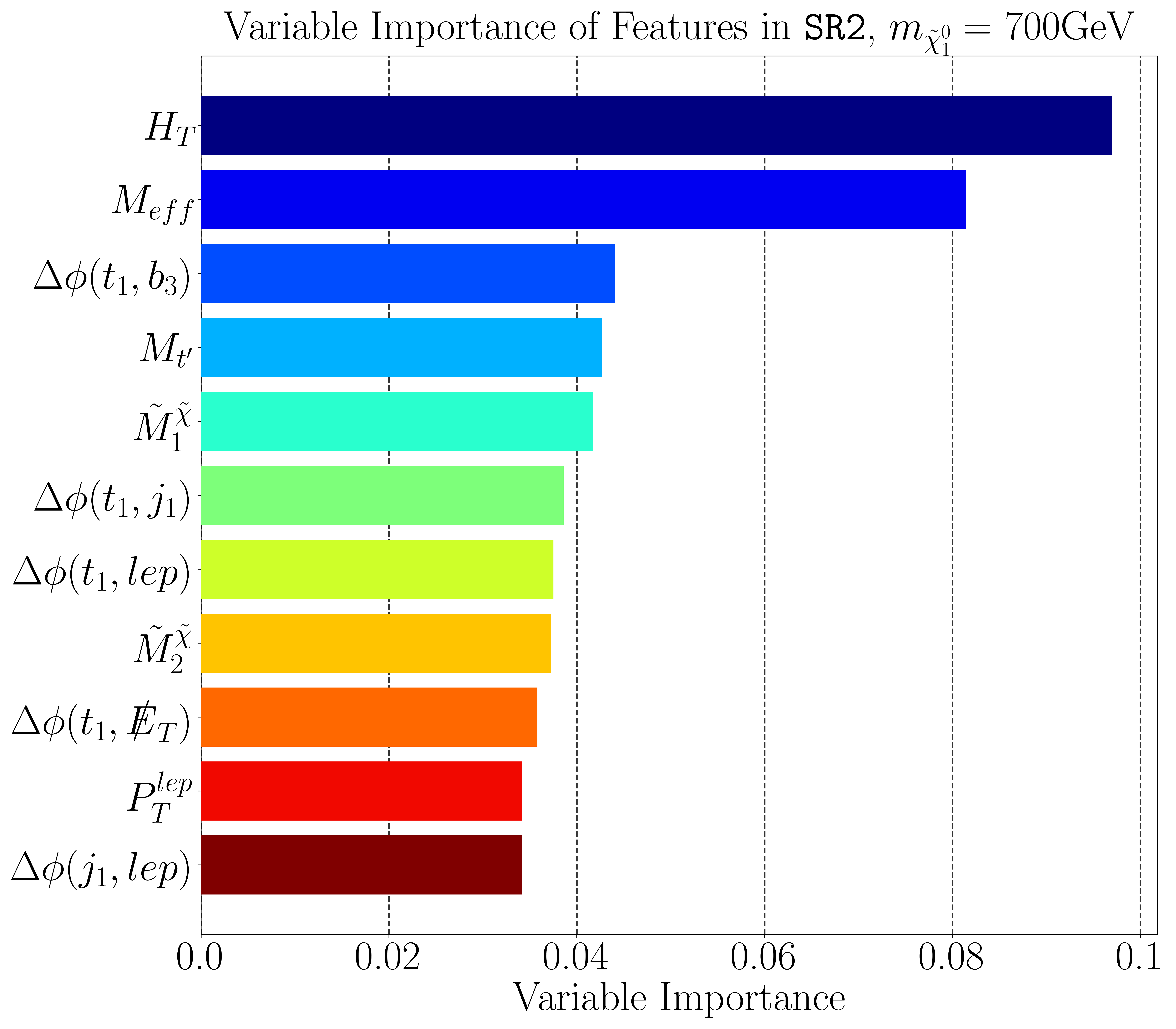}
	\caption{Variable Importance score of the kinematic variables as features of the BDTs trained for $m_{\tilde{\chi}}=700$ for \texttt{SR1} and \texttt{SR2}.} 
	\label{fig:var_imp}
\end{figure}

The variable importance is an indicator to showcase which features (kinematic variables) used in the BDTs are the most effective for the classification of signal and background. The top 10 most \textit{important} variables in the BDTs trained for $m_{\tilde{\chi}}=700$ GeV for both the signal regions are shown in Figure \ref{fig:var_imp}.

\newpage
\vspace{+2cm}
\section{Machine Learning Hyper-parameters}
\label{hyper_params}
We summarize the relevant hyperparameters of the BDT-based classifiers used for classifying the signal and background events in Table~\ref{table:bdtParametrs}.
\begin{table}[htb!]
	\centering
	\caption{\label{table:bdtParametrs} Summary of optimised BDT hyperparameters.}
	\begin{tabular}{ll}
		\toprule
		BDT hyperparameter & Optimised choice \\
		\midrule
		NTrees & 1000 \\ 
		MinNodeSize  & 5\% \\ 
		MaxDepth & 4 \\ 
		BoostType & AdaBoost\\ 
		AdaBoostBeta & 0.1  \\ 
		UseBaggedBoost & True \\ 
		BaggedSampleFraction & 0.5 \\
		SeparationType & GiniIndex \\ 
		nCuts & -1 \\ 
		\bottomrule
	\end{tabular} 
	
\end{table}

\newpage

\section{Sample yield for signal and background}
\label{yield}

\begin{table}[h!]
	\centering
	\renewcommand{\arraystretch}{1.1}
	\setlength{\tabcolsep}{9pt}
	\caption{ Table showcasing the total number of events in terms of signal yield and background yield at $\mathcal{L}=$ 3000 $fb^{-1}$ for $m_{\tilde{\chi}}$ = 500 GeV and $m_{\tilde{\chi}}$ = 700 GeV, for each BDT score bins along with the $Z_{exc}$ with the different values of background uncertainties.}
	\label{data_table}
	\begin{tabular}{||c||>{\raggedleft\arraybackslash}p{1.5cm}|>{\raggedleft\arraybackslash}p{1.5cm}|p{1.cm}|p{1.cm}|p{1.cm}|p{1.cm}||}
		\hline
		\textbf{BDT Score Bins}      & \textbf{Sig.} \textbf{Yield} & \textbf{Bkg.} \textbf{Yield}& $Z_{exc}$  $(0\% )$  & \textbf{$Z_{exc}$  $(1\% )$} & \textbf{$Z_{exc}$  $(3\% )$} & \textbf{$Z_{exc}$  $(5\% )$} \\ 
		\hline
		\hline
		\multicolumn{7}{||c||}{Benchmark Point at $m_{\tilde{\chi}}$ = 500 GeV} \\ 
		\hline
		\hline
		\textbf{\texttt{SR1}  (-1.0,-0.5)} & 2.90       & 13573.02    & 0.0249  & 0.0162  & 0.0068  & 0.0042  \\
		\hline
		\textbf{\texttt{SR1}  (-0.5,0.0)} & 726.06     & 246754.68   & 1.4595  & 0.2882  & 0.0978  & 0.0587  \\ 
		\hline
		\textbf{\texttt{SR1}  (0.0,0.2)} & 1156.60    & 86611.71    & 3.9040  & 1.2591  & 0.4404  & 0.2653  \\ 
		\hline
		\textbf{\texttt{SR1}  (0.2,0.4)} & 977.37     & 28211.23    & 5.7208  & 2.9475  & 1.1202  & 0.6803  \\ 
		\hline
		\textbf{\texttt{SR1}  (0.4,1.0)} & 69.27      & 842.83      & 2.2936  & 2.2587  & 1.7653  & 1.3237  \\ 
		\hline
		\textbf{\texttt{SR2}  (-1.0,-0.5)} & 0.00       & 7023.42     & 0.0000  & 0.0000  & 0.0000  & 0.0000  \\ 
		\hline
		\textbf{\texttt{SR2}  (-0.5,0.0)} & 332.60     & 1134027.89  & 0.3123  & 0.0292  & 0.0098  & 0.0059  \\ 
		\hline
		\textbf{\texttt{SR2}  (0.0,0.2)} & 499.48     & 184593.40   & 1.1610  & 0.2633  & 0.0898  & 0.0540  \\ 
		\hline
		\textbf{\texttt{SR2}  (0.2,1.0)} & 125.94     & 11302.55    & 1.1781  & 0.8094  & 0.3532  & 0.2182  \\ 
		
		\hline
		\multicolumn{3}{||c|}{\textbf{Combined Signal Significance}}  & 7.6284 & 4.0498 & 2.2053 & 1.5742 \\ 
		\hline
		\hline
		
		\multicolumn{7}{||c||}{Benchmark Point at $m_{\tilde{\chi}}$ = 700 GeV}\\
		\hline
		\hline
		\textbf{\texttt{SR1}  (-1.0,-0.5)} & 3.02       & 64571.20   & 0.0119  & 0.0043  & 0.0015  & 0.0009  \\
		\hline
		\textbf{\texttt{SR1}  (-0.5,0.0)} & 152.60     & 215384.00  & 0.3287  & 0.0692  & 0.0236  & 0.0142  \\ 
		\hline
		\textbf{\texttt{SR1}  (0.0,0.2)} & 283.83     & 61566.60   & 1.1413  & 0.4270  & 0.1521  & 0.0918  \\ 
		\hline
		\textbf{\texttt{SR1}  (0.2,0.4)} & 453.36     & 30999.40   & 2.5563  & 1.2663  & 0.4767  & 0.2892  \\ 
		\hline
		\textbf{\texttt{SR1}  (0.4,1.0)} & 145.36     & 3471.82    & 2.4168  & 2.1071  & 1.2001  & 0.7827  \\ 
		\hline
		\textbf{\texttt{SR2}  (-1.0,-0.5)} & 0.33       & 233401.00  & 0.0007  & 0.0001  & 0.0000  & 0.0000  \\ 
		\hline
		\textbf{\texttt{SR2}  (-0.5,0.0)} & 60.74      &901848.00  & 0.0640  & 0.0067  & 0.0022  & 0.0013  \\ 
		\hline
		\textbf{\texttt{SR2}  (0.0,0.2)} & 110.39     &165748.00  & 0.2711  & 0.0647  & 0.0221  & 0.0133  \\ 
		\hline
		\textbf{\texttt{SR2}  (0.2,1.0)} & 112.61     &35948.00   & 0.5930  & 0.2768  & 0.1027  & 0.0622  \\ 
		
		\hline
		\multicolumn{3}{||c|}{\textbf{Combined Signal Significance}} & 3.7704 & 2.5201 & 1.3279 & 0.8608 \\
		\hline
		
	\end{tabular}
	
\end{table}

\newpage
\bibliographystyle{JHEP}
\bibliography{v0}

\end{document}